\documentstyle[preprint,aps]{revtex}
\newcommand{\be}{\begin{equation}}
\newcommand{\ee}{\end{equation}}
\newcommand{\eq}{\begin{equation}}
\newcommand{\en}{\end{equation}}
\newcommand{\bc}{\begin{center}}
\newcommand{\ec}{\end{center}}

\newcommand{\lsim}{\raisebox{0.3mm}{\em $\, <$}
\hspace{-3.3mm} \raisebox{-1.8mm}{\em $\sim \,$}}

\begin{document}
\draft
\preprint{}
\title{Neutrino Flavor Mixing Constrained by Accelerator and
Reactor Experiments and Solar Neutrino Observation%
 \thanks{Work supported in part by Grant-in-Aid for Scientific Research of %
the Ministry of Education, Science and Culture \#0560355.}}
\author{Hisakazu Minakata
 \thanks{email: minakata@phys.metro-u.ac.jp}}
\address{\it Department of Physics, Tokyo Metropolitan University,\\
          1-1 Minami-Osawa, Hachioji, Tokyo 192-03 Japan}
\preprint{
\parbox{5cm}{
TMUP-HEL-9503\\
March 29, 1995\\
Revised: June 16, 1995\\
hep-ph/9504222\\
}}
\maketitle
\begin{abstract}

We derive the constraints imposed on neutrino masses and mixing angles
by performing a combined analysis of the data from the Los Alamos and
the other terrestrial neutrino oscillation experiments with the
assumption of the flavor-mixing solutions to the solar neutrino problem.
In a three-flavor mixing scheme which ignores the possibility of sterile
neutrinos, we obtain severe constraints on the pattern of masses and
flavor mixing of neutrinos. For example, we show that in the standard
Kobayashi-Maskawa type mixing matrix $s_{13}^2 \lsim 10^{-2}$
independent of the choice of the solar neutrino solutions. The constraint
from the double $\beta$ decay is also discussed.

\end{abstract}
\newpage

Recently the Liquid Scintillator Neutrino Detector(LSND) group in Los
Alamos has reported positive evidence for the neutrino oscillation events
$\bar{\nu_\mu} \rightarrow \bar{\nu_e}$\cite{NYT,Louis}. If confirmed by
continuing runs, it provides the first direct evidence for nonvanishing
masses and flavor mixings of neutrinos. It is certainly an urgent problem
to illuminate all possible consequences of this discovery\cite{PHKC}.

Before hearing this exciting report we have been speculating about the
neutrino masses and mixing with the help of information of astrophysical
neutrinos detected by the various underground detectors. Among them the
most popular one is the solar neutrino deficit which lasts more than 20
years \cite{Bahcall}, but still refuses resolution by the standard
solar model \cite{SSM}. Now the solar neutrino deficit is confirmed to
exist by the four different experiments, the chlorine\cite{Cleveland}, the
Kamiokande\cite{HS} and the two gallium experiments\cite{Abazov,Anselmann}.
A less popular but an equally important piece of information comes from
the atmospheric neutrino anomaly, a deficit in the ratio $(\nu_\mu + \bar
{\nu_\mu})/(\nu_e +\bar{\nu_e})$, which was first claimed to exist by
the Kamiokande group \cite{Hirata}, and subsequently supported by the
IMB \cite{Casper} and the Soudan 2 \cite{Goodman} experiments.
The new ``multi-GeV'' data sample reported by the Kamiokande group
\cite{Fukuda} provides strong support for the interpretation of
the atmospheric neutrino anomaly in terms of the neutrino oscillation.

In this paper we discuss possible implications of the LSND result by
performing a combined analysis of the data with that of solar neutrino
observation and the reactor experiments. Our analysis is based on the
three-flavor mixing scheme of neutrinos without introducing any
sterile species. We will recognize, as the analysis proceeds, that
it is crucial to include the information of the disappearance
experiments.  The most stringent limits achieved for $\nu_e$ and
$\nu_\mu$ channels are the ones reported by the Bugey \cite{Achkar} and
the CDHS \cite{CDHS} groups, respectively.

The most disturbing feature of this restricted framework is, of course,
that it cannot accomodate all of them, the LSND result, the solar neutrino
deficit, and the atmospheric neutrino anomaly. The reason is very simple;
three mass scales (more precisely, mass-squared differences) which are
involved in these three phenomena are too far apart from each other to be
accomodated in the three-generation scheme\cite{PHKC}. In the present paper
we therefore omit atmospheric neutrino data from the analysis. A combined
analysis of the alternative combination, LSND-Atmospheric neutrino,
is presented elsewhere \cite{mina1}.

To orient our analysis and to focus some important points we summarize below
the features of the LSND and the disappearance experiments, and the flavor
mixing solution to the solar neutrino problem.

\noindent
(1) The LSND experiment \cite{Louis,LSND,Hill}:

By using low energy $\bar{\nu_\mu}$ beam from stopped muons the LSND
group is able to perform the appearance experiment $\bar{\nu_\mu} \rightarrow
\bar{\nu_e}$ which can probe the neutrino oscillation parameters up to
$\Delta m^2 = 0.2$eV$^2$(0.5eV$^2$) for $\sin^2 2\theta = 0.01(0.001)$.
Such great sensitivity has been achieved by the intense pion beam from the
LAMPF accelerator and by the low enough energies of $\bar{\nu_\mu}$ beam.
After oscillating into $\bar{\nu_e}$, it produces positrons via the
reaction $\bar{\nu_e}p \rightarrow e^+ n$ which are measured in the energy
window 36MeV $<$ E$_e$ $<$ 60 MeV. The produced neutrons are captured by
protons through the reaction np $\rightarrow$ d$\gamma$, whose gamma rays
of energy 2.2 MeV can serve for delayed coincidence. The experimental
group may have observed excess in gamma ray-correlated events over the
estimated background.

\noindent
(2) The disappearance experiments \cite{Achkar,CDHS}:

This type of experiment measures the attenuation of neutrino beam from
a reactor or an accelerator. The intensity of $\bar{\nu_e}$ beam from
the reactor at Bugey is measured by the $^6$Li-loaded liquid scintillator
located at 15, 40 and 95m from the reactor \cite{Achkar}. From the
viewpoint of neutrino mixing the experiment measures
$1-P(\bar{\nu_e} \rightarrow \bar{\nu_e})$,
where $P(\bar{\nu_e} \rightarrow \bar{\nu_e})$ implies the survival
probability of electron antineutrinos. The resulting bound is rather
severe and may be summarized as $1-P(\bar{\nu_e} \rightarrow \bar{\nu_e})
\leq$ 5 \% including the statistical and the systematic uncertainties.
A comparable limit is achieved for $\nu_\mu$ channel by the CDHS
group \cite{CDHS} using $\nu_\mu$ beam from the CERN PS.

\noindent
(3) The flavor-mixing solutions to the solar neutrino problem:

The most popular and perhaps most appealing solution to the solar neutrino
problem is the one provided by the Mikhyev-Smirnov-Wolfenstein(MSW)
mechanism \cite{MSW}. It utilizes the (by now familiar) mechanism of
resonant enhancement of neutrino flavor conversion in solar matter.
At the moment there exist two options in this type of solution;
the small-angle (nonadiabatic), and the large-angle solutions. The
characteristic values of the neutrino mixing parameters are determined by,
for example, an extensive analysis by Hata and Langacker \cite{HL} and
are given as
$\Delta m^2 \simeq 6\times 10^{-6} \mbox{eV}^2, \quad
\sin^2 2\theta \simeq 7\times 10^{-3}$, and
$\Delta m^2 \simeq 9\times 10^{-6} \mbox{eV}^2, \quad
\sin^2 2\theta \simeq 0.6 $,
for the small-angle and the large-angle solutions, respectively.
Their analysis is done with the two-flavor mixing scheme, and
$\Delta m^2$ and $\theta$ indicate the mass-squared difference and
the mixing angle, respectively. We should mention that there still
exists the possibility of vacuum neutrino oscillation as a mechanism
for the solar neutrino deficit. This is the old solution \cite{Bahcall}
but one still alive \cite{BPW}.

One of the most important features in our combined analysis of the
LSND and the solar neutrino solution is that a huge mass hierarchy
is involved. (In fact, it is not really the hierarchy in mass but is
the hierarchy in the mass-difference.) One may classify the hierarchy
of neutrino masses into two types:
\begin{equation}
\mbox{(a) }\, m_1^2 \approx m_2^2 \gg m_3^2 \hskip 1cm
\mbox{(b) }\, m_3^2 \gg m_2^2 \approx m_1^2
\label{eqn:hierarchy}
\end{equation}
Here the symbols $\approx$ and $\gg$ imply the difference by
$\lsim 10^{-5}$eV$^2$ and $\sim$1-10eV$^2$, respectively.
In (\ref{eqn:hierarchy}) we have chosen the third state as the grossly
departed mass eigenstate. The other types of mass hierarchies which can
be obtained by permuting 1, 2, and 3 can be taken care of by an appropriate
choice of angles because they merely represent relabeling of the mass
eigenstates. Unlike the case of vacuum neutrino oscillation \cite{mina1}
the relative magnitude of the masses connected by $\approx$ does have
important meaning and will be discussed below.

We briefly review the neutrino oscillation with three flavors of
neutrinos. We define the neutrino mixing matrix $U$ which relates
the flavor eigenstate $\nu_\alpha (\alpha = e,\mu,\tau)$ and the
mass-eigenstate $\nu_i (i=1,2,3)$ in vacuum as $\nu_\alpha =
U_{\alpha i}\nu_i$. In this paper we assume CP invariance.
The evolution equation of the flavor eigenstate takes the form

\begin{equation}
i\frac{d}{dx}\nu_\alpha =
\frac{m_i^2}{2E}\delta_{ij}U_{\alpha i}(U^{-1})_{j\beta}\nu_\beta
+ \delta_{\alpha e}\delta_{e \beta}a_e(x)\nu_\beta,
\label{eqn:flavor}
\end{equation}
where $a_e(x)=\sqrt{2}G_F N_e(x)$ indicates the effect of matter
potential which affects only electron neutrinos. Here, $G_F, E$ and
$N_e$ denote, in order, the Fermi constant, the neutrino energy, and
the electron number density in the sun.

Without matter effect the equation (\ref{eqn:flavor}) can be easily
integrated to yield the oscillation probability
$\nu_\alpha \rightarrow \nu_\beta$ as

\begin{eqnarray}
P(\nu_\beta \rightarrow \nu_\alpha) &=& P (\bar{\nu_\beta} \rightarrow
\bar{\nu\_\alpha})\nonumber\\
        &=& \delta_{\alpha\beta}
-4\sum_{j>i} U_{\alpha i} U_{\beta i} U_{\alpha j} U_{\beta j}
\sin^2(\frac{\Delta m_{ij}^2 L}{4E}).
\label{eqn:probability}
\end{eqnarray}
The mass-squared difference $\Delta m_{ij}^2$ is defined as
$\Delta m_{ij}^2 \equiv m_j^2 - m_i^2  (j>i)$ in this paper. Notice
that when we discuss the neutrino evolution in matter the sign of
$\Delta m_{ij}^2$ does have physical meaning, unlike the case of vacuum
neutrino oscillation (\ref{eqn:probability}).

With matter effect the equation becomes complicated but it is
tractable because of the mass hierarchy involved in our analysis.
To make our discussion transparent we specify the form of the mixing
matrix in a form $U = U_{23}U_{13}U_{12}$,
\begin{equation}
U=\left[
\begin{array}{ccc}
	1 & 0 & 0\\
	0 & c_{23} & s_{23}\\
	0 &-s_{23} & c_{23}
\end{array}
\right]
\left[
\begin{array}{ccc}
	c_{13} & 0 & s_{13}\\
	0      & 1 & 0\\
	-s_{13}& 0 & c_{13}
\end{array}
\right]
\left[
\begin{array}{ccc}
	c_{12} & s_{12} & 0\\
	-s_{12}& c_{12} & 0\\
	0      & 0      & 1
\end{array}
\right],
\label{eqn:matrixU}
\end{equation}
where $U_{23}, U_{13},$ and $U_{12}$ denote the three matrices in
(\ref{eqn:matrixU}), in order, from left to right. Here $c_{ij}$
and $s_{ij}$ are the short-hand notations for $\cos \theta_{ij}$
and $\sin \theta_{ij}$, respectively. This is nothing but the standard
form of the Kobayashi-Maskawa matrix \cite{LEP}, which is now employed
as the neutrino mixing matrix. We note that CP invariance renders
the three angles real and they can all be made to lie in the first
quadrant by an appropriate redefinition of neutrino phases.

The definition of the mixing matrix (\ref{eqn:matrixU}) is convenient
in dealing with the mass patterns (\ref{eqn:hierarchy}) in which 1-2
level crossing is responsible for the solar neutrino deficit. If one
wants to discuss the other type of mass pattern which can be obtained
by permuting 1, 2, and 3 one may make a different choice of the $U$
matrix (i.e., redefinition of angles) convenient for them. For example,
$U = U_{13}U_{12}U_{23}$ for 2-3 level crossing. We note that one can
take  $\theta_{12} \leq \frac{\pi}{4}$ so that $\nu_e$ is close to
$\nu_1$ under the two-level crossing approximation. It does not hurt
the generality of our analysis because the constraints from the terrestrial
experiments to be discussed below do not involve $\theta_{12}$.

We examine the MSW effect in the three-flavor mixing scheme. We multiply,
following Kuo and Pantaleone \cite{KP},
$U_{13}^{-1}U_{23}^{-1}$ to the equation (\ref{eqn:flavor}) to obtain
the evolution equation for the modified neutrino basis
$\tilde{\nu_\beta}=(U_{13}^{-1}U_{23}^{-1})_{\beta\alpha}\nu_\alpha$:

\begin{equation}
i\frac{d}{dx}
\left[
\begin{array}{c}
\tilde{\nu_e}\\
\tilde{\nu_\mu}\\
\tilde{\nu_\tau}
\end{array}
\right]
=
\left[
\begin{array}{ccc}
-\Delta\cos 2\theta_{12}+c_{13}^2a_e & \Delta\sin 2\theta_{12} &
\frac{1}{2}\sin 2\theta_{13}a_e \\
\Delta\sin 2\theta_{12} & \Delta\cos 2\theta_{12} & 0\\
\frac{1}{2}\sin 2\theta_{13}a_e & 0 & \sum + s_{13}^2a_e\\
\end{array}
\right]
\left[
\begin{array}{c}
\tilde{\nu_e}\\
\tilde{\nu_\mu}\\
\tilde{\nu_\tau}
\end{array}
\right],
\label{eqn:Kuo}
\end{equation}
where we have used simplified notation
$\Delta \equiv \frac{\Delta m_{12}^2}{4E}$ and
$\sum \equiv \frac{1}{2}(\Delta m_{13}^2+\Delta m_{23}^2)$.
Note that the sign of $\Delta$ has physical significance.

{}From (\ref{eqn:Kuo}) one realizes that the effective two-level crossing
approximation is justified unless $\sin 2\theta_{13}a_e$ is
extraordinarily large compared with other elements, which is not
the case in our problem. Moreover, an evaluation of the perturbative
corrections to the energy eigenvalues due to this off-diagonal term
reveals that they are of the order of $-(\sin 2\theta_{13} a_e)^2/|\sum|$
\cite{KP} which is negligible for the mass hierarchy
$\Delta m_{13}^2 \approx \Delta m_{23}^2 \gg \Delta m_{12}^2$.
Notice that this is true for both of the types-a and -b mass patterns
given in (\ref{eqn:hierarchy}). Also it can be shown that the
correction to the difference between two energy eigenvalues in matter
vanishes at the resonance point. Therefore, the off-diagonal terms
affect the discussion of the adiabaticity condition in the effective
two-level problem only through higher-order corrections.

Having established the validity of the effective two-level approximation
we proceed to the combined analysis. We first derive the approximate
formulas for the terrestrial neutrino experiments, taking into account
the mass hierarchy and the experimental parameters. With mass hierarchy
(\ref{eqn:hierarchy}) the oscillation probability consists of two terms,
one involving large $\Delta m_{13}^2$ and other small $\Delta m_{12}^2$.
The coefficient of the former term have a simplified expression due to
the orthogonality of the mixing matrix \cite{mina1}. The latter term is
smaller by factors of $(\Delta m_{12}^2 / \Delta m_{13}^2)^2 \leq 10^{-10}$
compared with the former. So it can be safely ignored.

In certain cases, the argument of the sine function with larger mass
difference takes the large values, e. g., 10-100 for
$\Delta m_{13}^2 = 1-10$eV$^2$ with E$=4$MeV and L$=40$m, the typical
parameters in the Bugey experiment. Therefore, it can be averaged
to be $\frac{1}{2}$ and we obtain, as the formula for the Bugey
disappearance experiment,

\begin{equation}
1-P(\bar{\nu_e} \rightarrow \bar{\nu_e}) = 2c_{13}^2s_{13}^2.
\label{eqn:Bugey}
\end{equation}
For the LSND experiments the arguments of sine factors with
$\Delta m_{13}^2$ and $\Delta m_{23}^2$ are of order unity and so we
cannot average. The oscillation probability in the LSND experiment
is thus

\begin{equation}
P(\bar{\nu_\mu} \rightarrow \bar{\nu_e}) =
4s_{23}^2c_{13}^2s_{13}^2 \sin^2 (\frac{\Delta m_{13}^2 L}{4E}).
\label{eqn:LSND}
\end{equation}
Similarly, the oscillation probabilities of $\nu_\mu \rightarrow \nu_\mu$
and $\nu_\mu \rightarrow \nu_\tau$ channels take the forms

\begin{equation}
1-P(\nu_\mu \rightarrow \nu_\mu) = 4s_{23}^2c_{13}^2
(1-s_{23}^2c_{13}^2)\sin^2 (\frac{\Delta m_{13}^2 L}{4E})
\label{eqn:CDHS},
\end{equation}
\begin{equation}
P(\nu_\mu \rightarrow \nu_\tau) =
4c_{23}^2s_{23}^2c_{13}^4\sin^2(\frac{\Delta m_{13}^2 L}{4E}).
\label{eqn:E531}
\end{equation}

{}From the data of the Bugey and the LSND experiments described earlier
we require

\begin{eqnarray}
\label{eqn:BL1}
c_{13}^2s_{13}^2 \leq 2.5 \times 10^{-2} \equiv \delta \\
\label{eqn:BL2}
s_{23}^2c_{13}^2s_{13}^2 \equiv \epsilon \lsim 10^{-3}
\end{eqnarray}
where we have defined $\delta$ and $\epsilon$ such that $2\delta$ and
$4\epsilon$ imply the attenuation of $\bar{\nu_e}$ in the Bugey and
the ``rate'' in the LSND experiments, respectively. Since the latter
quantity is still subject to the uncertainty \cite{LSND,Hill}, we take
a conservative attitude and assume that $\epsilon$ is less than or equal
to $\sim 10^{-3}$. From (\ref{eqn:BL1}) and (\ref{eqn:BL2}) it follows
that $\theta_{13}$ must be either small or close to $\frac{\pi}{2}$,

\begin{equation}
\label{eqn:BL3}
\epsilon \leq
\left(
\begin{array}{c}
s_{13}^2\\
c_{13}^2
\end{array}
\right)
\leq 2.6 \times 10^{-2} \simeq \delta
\end{equation}

The angle $s_{23}$ is severely constrained by the the $\nu_\mu$
disappearance experiment. The CDHS group \cite{CDHS} obtained the
constraint on $\sin^2 2\theta$ (in two-flavor scheme),
$\sin^2 2\theta \lsim 0.1$ in the mass range
$1 eV^2 \leq \Delta m_{13}^2 \leq 10 eV^2$, where the constraint from
the CDHS experiment is most stringent. Using this with (\ref{eqn:CDHS})
we obtain the bound on $s_{23}^2$ as

\begin{equation}
\label{eqn:S23CDHS}
\left(
\begin{array}{c}
s_{23}^2c_{13}^2\\
1-s_{23}^2c_{13}^2
\end{array}
\right)
\lsim 2.5 \times 10^{-2} \simeq \delta.
\end{equation}
In large-$\Delta m_{13}^2$ region a better bound can be obtained from
$\nu_\mu \rightarrow \nu_\tau$ oscillation experiment. Using the data
of Fermilab E531 experiment \cite{E531} with (\ref{eqn:E531})
we obtain an approximate expression of the $\Delta m^2$-dependent bound
in the same mass range:

\begin{equation}
\label{eqn:S23E531}
c_{23}^2s_{23}^2c_{13}^4 \lsim 0.25 \times
\left(
\frac{\Delta m_{13}^2}{1 \rm{eV}^2}\right)^{-2},
\end{equation}

We will show below that the solar neutrino solutions force us to choose the
small-$s_{13}$ solution out of (\ref{eqn:BL3}). Using this information
apriori with (\ref{eqn:S23CDHS}) and (\ref{eqn:S23E531}) we obtain the
bound on $s_{23}^2$,

\begin{equation}
\label{eqn:S23}
\left(
\begin{array}{c}
s_{23}^2\\
c_{23}^2
\end{array}
\right)
\lsim 2.5 \times 10^{-2} \times
\left\{
\begin{array}{ll}
1 & (1 \rm{eV}^2 \leq \Delta m_{13}^2 \leq 3.3 \rm{eV}^2) \\
\left( \frac{\Delta m_{13}^2}{3.3 \rm{eV}^2} \right)^{-2} &
(3.3 \rm{eV}^2 \leq \Delta m_{13}^2 \leq 10 \rm{eV}^2)
\end{array}
\right.
\end{equation}

Now let us turn to the constraint implied by the solar neutrino solutions.
Using the local two-level crossing approximation established above
$\Delta m_{12}^2$ and $s_{12}^2$ are determined to be

\begin{equation}
\label{eqn:deltam1}
\Delta m_{12}^2 \simeq 6\times 10^{-6}\mbox{eV}^2, \hskip 0.5cm
s_{12}^2 \simeq 1.75 \times 10^{-3}, \hskip 1cm (\mbox{small-angle})
\end{equation}
and
\begin{equation}
\label{eqn:deltam2}
\Delta m_{12}^2 \simeq 9\times 10^{-6}\mbox{eV}^2, \hskip 0.5cm
s_{12}^2 \simeq 0.184, \hskip 1cm (\mbox{large-angle})
\end{equation}
for the small-angle and the large-angle MSW solutions, respectively.
Of course, the mass-squared difference $\Delta m_{12}^2$ must be
positive in order that the MSW mechanism acts for neutrinos, not for
antineutrinos. This implies $m_{2}^2 > m_{1}^2$ between almost
degenerate neutrino states in (\ref{eqn:hierarchy}).

So far we have discussed the constraints from the terrestrial experiments
and that from the MSW solar neutrino solutions separately (except what
we did for $s_{23}$). While we have obtained highly nontrivial constraints
(\ref{eqn:BL3})-(\ref{eqn:S23E531}), it would be more interesting if there
arise further restrictions by considering these three experimental
requirements simultaneously. We show that this in fact occurs.

We demonstrate that the large-$s_{13}$ solution (i. e.,
$c_{13}^2 \leq 2.6\times 10^{-2}$) is not acceptable as a solution to
the solar neutrino problem. To show this we note the relationship
between the flavor basis $\nu_\alpha$ and the basis $\tilde{\nu_\alpha}$
introduced as a convenient basis for describing the effectively local
two-level resonance in neutrino evolution in matter. It is

\begin{equation}
\left[
\begin{array}{c}
\tilde{\nu_e}\\
\tilde{\nu_\mu}\\
\tilde{\nu_\tau}
\end{array}
\right]
=
\left[
\begin{array}{ccc}
c_{13} & -s_{23}s_{13} & -c_{23}s_{13}\\
0 & c_{23} & -s_{23}\\
s_{13} & s_{23}c_{13} & c_{23}c_{13}
\end{array}
\right]
\left[
\begin{array}{c}
\nu_e\\
\nu_\mu\\
\nu_\tau
\end{array}
\right].
\label{eqn:matrix}
\end{equation}
It we take the large-$s_{13}$ solution the initial condition
$[\nu_e \quad \nu_\mu \quad \nu_\tau] \simeq [1, 0, 0]$ at the solar
core is translated into the initial condition
$[\tilde{\nu_e} \quad \tilde{\nu_\mu} \quad \tilde{\nu_\tau}] =
[\sqrt{\delta}, 0, 1]$.
Since $\tilde{\nu_\tau}$ effectively decouples with $2\times2$
submatrix which has resonance, as can be seen in (\ref{eqn:Kuo}),
it experiences no significant change in the solar interior. Therefore,
$\nu_e$'s which departed the solar core just leave the sun with no
appreciable attenuation in their flux.\footnote{One can reach the same
conclusion by using the local two-level crossing representation of 1-3
channel\cite{KP}. In this case ``$\tilde{\nu_\tau}$'' does not experience
a level crossing not because it decouples but because the resonance
condition cannot be met due to the mass hierarchy
$\Delta m_{12}^2 \ll \Delta m_{13}^2 \approx \Delta m_{23}^2$
with which we are working.}
So the large-$s_{13}$ solution cannot explain the solar neutrino deficit.
This conclusion is valid for both of the small-angle and the large-angle
MSW solutions, because the relation (\ref{eqn:matrix}) is independent
of $s_{12}$. Thus, we are left with the small-$s_{13}$ solution with
the additional constraints (\ref{eqn:S23}) together with either
(\ref{eqn:deltam1}), or (\ref{eqn:deltam2}), depending upon the
small-angle and the large-angle MSW solutions, respectively.

Now we turn to the discussion of the solar neutrino solution based on
the vacuum neutrino oscillations. Because of the mass hierarchy involved,
1-10eV$^2 \simeq \Delta m_{13}^2 \approx \Delta m_{23}^2 \gg
\Delta m_{12}^2 \simeq 10^{-10}$eV$^2$,
the relevant oscillation probability can be written as

\begin{equation}
P(\nu_e \rightarrow \nu_e) =
1-2c_{13}^2s_{13}^2-4c_{12}^2s_{12}^2c_{13}^4\sin^2
(\frac{\Delta m_{12}^2L}{4E}),
\label{eqn:osci}
\end{equation}
where the sine-squared terms with $\Delta m_{13}^2$ and
$\Delta m_{23}^2$ are averaged. It is an excellent approximation because
the argument of sine takes a very large value,
$\simeq 2\times 10^{10} (\Delta m^2 / 1 \mbox{eV}^2)(L/1\mbox{AU})
(E/10\mbox{MeV})^{-1}$.
In view of (\ref{eqn:osci}) and noticing the Bugey constraint (\ref{eqn:BL1})
we realize that the only possible way of having solar neutrino deficit
of $\sim$50\% level is to have an effective two-flavor description
of $P(\nu_e \rightarrow \nu_e)$. Namely, we have to demand

\begin{equation}
\begin{array}{c}
 c_{12}^2s_{12}^2 \simeq \frac{1}{4} (=\mbox{maximum value}), \\
 c_{13}^2 \simeq 1,
\end{array}
\label{eqn:gross}
\end{equation}
to have a gross deficit. Under the condition (\ref{eqn:gross}) the vacuum
neutrino oscillation (\ref{eqn:osci}) will provide an acceptable solution
to the solar neutrino problem (with possible slight changes in mixing
parameters) as shown in \cite{Kras}. Thus, the combined analysis of
the LSND and the Bugey experiments with the vacuum mixing solution of
the solar neutrino problem again prefers the small-$s_{13}$ solution.

An additional constraint arises in the case of Majorana neutrinos.
The quantity

\begin{eqnarray}
<m_{\nu e}> &=& \sum_{j=1}^{3} \eta_j |U_{ej}|^2 m_j\nonumber\\
        &=& \eta_1 c_{12}^2c_{13}^2 m_1
+ \eta_2 s_{12}^2c_{13}^2 m_2
+ \eta_3 s_{13}^2 m_3
\label{eqn:beta}
\end{eqnarray}
is constrained to be less than $\sim$1 eV by the non-observation of
the neutrinoless double $\beta$ decay in various experiments \cite{Moe}.
Notice that we are working with the representation in which the mixing
matrix is real under the assumption of CP invariance, and
$\eta_j =  \pm 1$ in (\ref{eqn:beta}) is the CP phase.

The constraint from the double $\beta$ decay acts differently for
the type-a and the type-b mass hierarchies in (\ref{eqn:hierarchy}).
In the type-a case there is a chance for cancellation between nearly
degenerate two masses, but there is no chance in the type-b case
because the heavy mass is carried by a unique mass eigenstate.

We, however, encounter a new situation in the consistent solutions
obtained in our combined analysis. For the type-b mass hierarchy the
double $\beta$ decay constraint is automatically satisfied by the
small-$s_{13}^2$. On the other hand, the constraint for the type-a mass
hierarchy has nontrivial consequences. In the case of small-angle
MSW solution, the cancellation between 1 and 2 mass eigenstates is
hopeless because $s_{12}^2$ is too tiny, $\simeq 10^{-3}$. In the case
of large-angle MSW solution the situation is better but we still obtain
$<m_{\nu e}> \simeq$ 1.7 eV for $m_1 = m_2$ = 2.4 eV, which is larger by
factor of $\sim$ 2 than the experimental bound \cite{Moe}. In the case
of vacuum oscillation solution, there exists better chance for cancellation
because $s_{12}^2$ is large. The double $\beta$ constraint can be
cleared with a mild condition $33^{\circ} \leq \theta_{12} \leq 57^{\circ}$.
Thus, the double $\beta$ constraint prefers the vacuum oscillation
solution for the type-a mass hierarchy, while it is automatically satisfied
for the type-b mass pattern.

In this paper we have discussed the constraints imposed on neutrino
masses and mixings when we demand the consistency with the LSND and
the other terrestrial neutrino experiments and the flavor mixing solutions
to the solar neutrino problem. Independent of the choice of the three
solar neutrino solutions, the small- and the large-angle MSW, and the
vacuum oscillation solutions, $s_{13}^2$ is constrained to be small,
$s_{13}^2 \lsim \delta \sim 10^{-2}$, for different reasons in the MSW
and in the vacuum solutions, respectively.  The angle $s_{23}^2$ is
subject to the constraint (\ref{eqn:S23}) which says that it is either
small, $\lsim 10^{-2}$ or large, $\sim 1$. On the other hand, the value
of $s_{12}^2$ depends upon the solar neutrino solutions;
$s_{12}^2 \simeq 1.8 \times 10^{-3}, 0.18, 0.5$ for the small-angle
MSW, the large-angle MSW, and the vacuum oscillation solutions,
respectively.

The physical interpretation of the solution is clearest in the
small-angle MSW solution. The $\nu_1$ state is approximately
identical with ``$\nu_e$'', and $\nu_2$ and $\nu_3$ are in general
mixtures of ``$\nu_\mu$''and ``$\nu_\tau$''. Therefore, the type-a
contains an inverted mass hierarchy and the type-b implies a normal
mass pattern. Likewise one can obtain analogous physical
interpretations in other solutions but the $\nu_1$ states is less
pure with larger mixing angles.

Finally, a few remarks are in order:

\noindent
(1) The present analysis is less powerful in constraining the absolute
values of masses than the relative masses and the mixing angles,
a general feature noticed in \cite{mina1}. The constraints obtained in
this paper do allow, for Dirac neutrinos, the type-b solution with e.g.,
$m_1 =$ 6 eV, $m_2 =$ 6 eV, and $m_3 =$ 6.5 eV, which is consistent
with the direct mass measurement \cite{Otten} and the cosmological
constraints \cite{Kolb}. This extreme choice would provide the possibility
that the light neutrinos could fill the entire part of the missing mass
in the universe, but with possible troubles with galaxy formation
\cite{Kolb,PHKC}.

\noindent
(2) One must be careful in comparing the resulting constraints obtained
in this paper with that of Ref. \cite{mina1} not only because they stand
on entirely different basis but also because the definition of the angles
are different.  This complication arises due to the fact that we are
working with the different definitions of the mass eigenstates here
and in Ref. \cite{mina1}. We are planning to present a unified and
more transparent description in the future \cite{mina2}.

I thank Hiroshi Nunokawa for discussions, Eligio Lisi and the referee
of Ref. \cite{mina1} for calling my attention to various terrestrial
experiments which have not addressed in the earlier version of this paper.
This work is supported in part by Grant-in-Aid for Sciectific Research
of the Ministry of Education, Science and Culture, \#0560355.

Note Added: After submitting the earlier version of this paper there
appeared two reports from the LSND group \cite{LSND,Hill} with mutually
conflicting conclusions.

\end{document}